\documentclass[a4paper,11pt]{article}
\usepackage{aaskaiid}

\title{VLBI-Enabled Precision Localization of Continuous Gravitational Wave Sources with Pulsar Timing Arrays in the SKA Era}
\ShortTitle{VLBI-Enabled Localization of Continuous GW Sources}

\author[1]{Keitaro Takahashi}
\ShortName{Takahashi et al.} 
\author[2]{Takuya Akahori}
\author[3]{Kenta Fujisawa}
\author[4]{Hiroshi Imai}
\author[5]{Hajime Kita}
\author[6]{Hideyuki Kobayashi}
\author[7]{Hiroaki Misawa}
\author[8]{Kotaro Niinuma}
\author[9]{Tomoaki Oyama}
\author[10]{Kazuhiro Takefuji}
\author[7]{Fuminori Tsuchiya}

\affiliation[1]{Faculty of Advanced Science and Technology, Kumamoto University, 2-39-1 Kurokami, Kumamoto 860-8555, Japan }
\emailAdd{keitaro@kumamoto-u.ac.jp}
\affiliation[2]{Mizusawa VLBI Observatory, National Astronomical Observatory of Japan, 2-21-1 Mitaka, Tokyo 181-8588, Japan}
\affiliation[3]{Research Institute for Time Studies, Yamaguchi University, Yoshida 1677-1, Yamaguchi-city, Yamaguchi, 753-8511, Japan}
\affiliation[4]{Amanogawa Galaxy Astronomy Research Center, Graduate School of Science and Engineering, Kagoshima University, 1-21-30 Korimoto, Kagoshima 890-0065, Japan}
\affiliation[5]{Faculty of Engineering, Tohoku Institute of Technology, 35-1 Yagiyama-Kasumicho, Thaihaku-ku, Sendai-shi, Miyagi, 982-8577 Japan}
\affiliation[6]{Center of Radio Astronomy and Engineering, National Astronomical Research Institute of Thailand, 260 Moo 4, Don Kaeo Subdistrict, Mae Rim District, Chiang Mai Province 50180, Thailand}
\affiliation[7]{Planetary Plasma \& Atmospheric Reseaerch Center, Graduate School of Science, Tohoku University, 6-3 Aramaki Aza Aoba, Aoba-ku, Sendai, 980-8578, Japan}
\affiliation[8]{Graduate School of Sciences and Technology for Innovation, Yamaguchi University, Yoshida 1677-1, Yamaguchi, 753-8512, Japan}
\affiliation[9]{Mizusawa VLBI Observatory, National Astronomical Observatory of Japan, 2-12 Hoshiga-oka, Mizusawa, Oshu-shi, Iwate 023-0861, Japan}
\affiliation[10]{Japan Aerospace Exploration Agency, Usuda Deep Space Center, 1831-6 Oomagari Kamiodagiri, Saku-City, Nagano, 384-0306, Japan}

\abstract{Pulsar timing arrays (PTAs) are opening the nanohertz gravitational-wave (GW) band by timing millisecond pulsars (MSPs) to target signals from supermassive black hole binaries (SMBHBs). Beyond evidence for a stochastic background, a central SKA-era objective is detecting individual continuous-wave (CW) sources. The scientific payoff hinges on localization: conventional PTA searches yield uncertainties of tens–hundreds of deg$^2$, too large to identify a unique host, obtain a redshift, infer intrinsic masses, or pursue electromagnetic counterparts. This limitation is chiefly geometric: the CW response includes Earth and pulsar terms, and poorly known pulsar distances make the pulsar-term phase a free parameter that degrades triangulation. If distances to a few MSPs are known to better than a GW wavelength ($\sim$1 pc), these phases are fixed and localization improves by orders of magnitude. Simulations indicate that with sub-parsec distances for a handful of nearby MSPs, the uncertainty can shrink to $\sim 10^{-3}$\,deg$^2$ (arcminute scale), enabling unique host association and multi-messenger follow-up. Achieving such distances requires $\sim 10~\mu$as parallaxes for MSPs within a few hundred parsecs, a precision now approached with Very Long Baseline Interferometry (VLBI) and expected to become practical with phased-array SKA1-Mid operating as a sensitive VLBI element. SKA1’s multi-beam, multi-calibrator astrometry will provide the independent distance priors needed for PTAs to localize nanohertz GW sources and measure SMBHB parameters and environments. We assess VLBI’s role in PTA CW searches and propose a concrete SKA1-Mid observing strategy for nearby MSPs to deliver the required sub-parsec distances.}

\begin{document}
\maketitle

\section{Introduction}

Pulsar timing arrays (PTAs) enable the detection of nanohertz-frequency gravitational waves (GWs) by monitoring pulse times of arrival from ensembles of millisecond pulsars (MSPs). A passing GW induces correlated perturbations (timing residuals) in the measured arrival times across the array. The dominant expected sources of such signals are supermassive black hole binaries (SMBHBs), formed during galaxy mergers and evolving over Gyr timescales. These systems are predicted to generate an approximately stochastic GW background arising from the superposition of many unresolved binaries, and continuous, quasi-monochromatic GW emission from individually bright binaries that are either nearby or especially massive. In recent years, indications of a stochastic GW background have emerged from ongoing PTA experiments \citep{2023ApJ...951L...8A,2023A&A...678A..50E,2023ApJ...951L...6R,2023RAA....23g5024X,2024ApJ...966..105A,2025MNRAS.536.1489M}, and a definitive detection is anticipated in the near future.

In the era of the Square Kilometre Array (SKA), PTAs are anticipated not only to refine measurements of the stochastic GW background, but also to detect individual continuous-wave (CW) sources \citep{2015aska.confE..37J}. The critical next objective is astrophysical localization: determining the sky position of a single SMBHB with sufficient precision to identify its host galaxy and to enable electromagnetic follow-up. Once the host galaxy is identified, one can obtain its redshift and basic properties (e.g.\ morphology, merger state, and nuclear activity). Combined with the observed GW strain amplitude and frequency, this enables inference of binary parameters such as orbital separation and chirp mass, and facilitates investigation of the role of gas, stellar scattering, and dynamical friction in binary evolution. This represents the onset of true multi-messenger nanohertz GW astronomy: the transition from the detection to the physical characterization of a specific SMBHB in a specific galaxy.

At present, source localization is the principal limitation. Standard PTA searches for a single CW source typically yield sky localization regions with areas of tens to hundreds of square degrees \citep{2016MNRAS.461.1317Z,2018MNRAS.477.5447G}. Such regions may contain hundreds to thousands of galaxies out to moderate redshift ($z \lesssim 0.2$), preventing an unambiguous association between the GW source and a unique host. In effect, the PTA is capable of detecting the binary but not of indicating its precise position. For host identification and targeted multi-wavelength follow-up, localization at the arcminute level is required, corresponding to areas of order $\sim 10^{-3}$ deg$^2$, rather than $\sim 10^2$ deg$^2$. At this scale, the number of plausible massive host galaxies within the localization region is reduced to order unity.

The origin of the poor localization in conventional PTA analyses can be understood in terms of the PTA response to a CW signal. The GW imprint in the timing residuals includes two components. The "Earth term" represents the GW signal at the Solar System Barycenter. The "pulsar term’’ represents the same GW signal at the pulsar, delayed by the light travel time from the pulsar to Earth. The pulsar term carries a phase that depends sensitively on the pulsar’s distance. If a pulsar’s distance is not well constrained, the pulsar term acts as an effectively free phase parameter for that pulsar. These unconstrained phases absorb information that would otherwise contribute to geometrical triangulation of the source position, thereby severely degrading localization.

\cite{PhysRevD.108.123535} showed that this degeneracy can be broken if distances to a subset of pulsars are known with sufficiently high precision. In this regime, the phases of the pulsar terms become fixed rather than free, and the addition of only a few well-characterized MSPs can improve localization by orders of magnitude. ``Sufficiently high precision'' in this context means that the pulsar distance is known to within a fraction of a GW wavelength. For nanohertz GWs ($f \sim 10^{-8}$ Hz), the corresponding wavelength is of order 1 pc. Consequently, to achieve arcminute-scale localization, distances to several PTA MSPs must be determined with uncertainties $\lesssim 1$ pc. When such priors are included, simulations indicate that the sky area for a strong CW source can be reduced from tens to hundreds of deg$^2$ to $\sim 10^{-3}$ deg$^2$ in favorable geometries \citep{2026PhRvD.113b2001K}. An area of $\sim 10^{-3}$ deg$^2$ corresponds to a few arcminutes on a side and is typically small enough to isolate a unique host galaxy.

Achieving sub-parsec distance precision requires micro-arcsecond-level astrometry. Very long baseline interferometry (VLBI) provides a direct geometric measurement of trigonometric parallax by tracking the absolute position of the pulsar on the sky with respect to background calibrators over the course of Earth's orbit. Recent VLBI programs have demonstrated parallax precisions at the $\sim 10~\mu$as level for nearby MSPs \citep{2013ApJ...770..145D,2016ApJ...828....8D,2019ApJ...875..100D,2023MNRAS.519.4982D}, which for MSPs within a few hundred parsecs corresponds to absolute distance uncertainties at or below $\sim 1$ pc. This makes VLBI-derived distances suitable for use as independent external priors in PTA continuous-wave analyses. In contrast, timing-parallax information is inferred from the same long-term timing data used in the PTA analysis itself and therefore does not constitute an independent external constraint. In a self-consistent continuous-wave analysis, it is already incorporated in the joint inference over the timing model and GW parameters, whereas VLBI astrometry can be incorporated as an external distance prior.

The emerging picture is therefore as follows. PTAs are in principle already capable of detecting CW sources, but by themselves they generally cannot localize these sources with sufficient precision to enable host-galaxy identification. VLBI astrometry provides precisely the external distance priors required to lift that degeneracy. Once PTA localization reaches the arcminute scale (i.e.\ $\sim 10^{-3}$ deg$^2$), it becomes possible to associate an individual nanohertz GW source with a specific galaxy, obtain the redshift of that system, search for active galactic nucleus activity or other electromagnetic counterparts, and infer binary parameters in physical units rather than in purely strain-based units. Such a capability would directly constrain the assembly and growth of supermassive black holes through mergers and the physical mechanisms by which SMBHBs harden in galactic nuclei. Sub-parsec MSP distance measurements are thus not merely auxiliary data products; they are central to enabling precision nanohertz GW astrophysics in the SKA era.

\section{Status of VLBI Astrometry of Millisecond Pulsars}

VLBI astrometry determines a pulsar’s absolute position on the sky at multiple epochs and fits for both proper motion and annual parallax. For MSPs, which are intrinsically faint radio sources (often a few mJy at $\sim$1–2 GHz) and have millisecond spin periods, such measurements were historically difficult. Over the past decade, however, dedicated VLBI programs have demonstrated that astrometric precisions of a few tens of micro-arcseconds can be obtained systematically \citep{2013ApJ...770..145D,2016ApJ...828....8D,2019ApJ...875..100D,2023MNRAS.519.4982D}. This level of angular precision corresponds directly to sub-parsec distance precision for nearby PTA MSPs, matching the localization requirements discussed above.

Several methodological advances have enabled this progress:

\textbf{Pulsar gating.} During VLBI correlation, the correlator can be configured to integrate only during the on-pulse portion of the pulsar’s spin phase. Because MSPs emit in a narrow duty cycle, typically $10\%$, such gating improves the effective signal-to-noise ratio by factors of several. This technique allows otherwise marginal detections of mJy-class MSPs to become usable astrometric detections at GHz frequencies \citep{2013ApJ...770..145D}. Gating is therefore essential for high-precision astrometry of faint, fast-spinning pulsars.

\textbf{Phase referencing and in-beam calibration.} High-precision VLBI astrometry is differential: the pulsar’s position is measured relative to a compact reference source. Conventional fast-switching phase referencing alternates between the pulsar and a nearby calibrator to track atmospheric and instrumental phase variations. An even more effective approach is in-beam phase referencing, in which the calibrator lies within (or close to) the pulsar’s primary beam and can be observed effectively simultaneously. Using multiple nearby calibrators allows reconstruction of a two-dimensional phase screen across the field, suppressing residual spatial phase gradients \citep{2016ApJ...828....8D,2019ApJ...875..100D,2023MNRAS.519.4982D}. This multi-calibrator strategy is what enables $\sim 10$–$50~\mu$as relative astrometry.

\textbf{Wide bandwidth and multi-epoch sampling.} Modern VLBI backends provide bandwidths of hundreds of MHz, increasing sensitivity and assisting in the separation of dispersive (ionospheric) and non-dispersive (tropospheric and geometric) delays. Observations are repeated over $\sim$6–10 epochs spanning $\sim$1–2 years to sample the full parallax ellipse in both right ascension and declination. This cadence constrains both the annual parallax and the secular proper motion.

These techniques have yielded several notable results. For PSR~J2222$-$0137, VLBI astrometry measured a parallax of $3.742^{+0.013}_{-0.016}$ mas, corresponding to a distance of $267.3^{+1.2}_{-0.9}$ pc, i.e.\ a fractional distance uncertainty of order $0.4\%$ \citep{2013ApJ...770..145D}. This implies sub-parsec absolute distance accuracy at $\sim 270$ pc. Fixing the distance in the pulsar timing model then enabled a precise detection of Shapiro delay and a tight constraint on orbital inclination and even on the longitude of ascending node. This illustrates the synergy between VLBI astrometry and high-precision timing: VLBI establishes the geometric framework (distance, position, proper motion), thereby allowing timing analyses to probe relativistic binary dynamics.

\cite{2016ApJ...828....8D} extended these methods to MSPs J1022+1001 and J2145$-$0750, obtaining distances of order 600–700 pc with uncertainties of $\sim$10–20 pc. These model-independent distances allowed re-analysis of optical photometry of the white dwarf companions, yielding mass estimates of $\sim 0.85 M_\odot$ and constraining cooling ages. Moreover, comparisons between VLBI-based astrometry and PTA timing solutions revealed discrepancies at the $\sim 5\sigma$ level in parallax and proper motion for some MSPs. Such differences were traced to systematic effects in timing models (for example, incomplete solar wind modeling), underscoring the robustness of VLBI astrometry as an external geometric reference for PTA pulsars.

\cite{2019ApJ...875..100D} expanded this approach to 57 pulsars in the PSR$\pi$ program, approximately doubling the number of pulsars with $\gtrsim 5\sigma$ trigonometric distance constraints. Typical parallax uncertainties in that survey were $\sim 45~\mu$as, with the best cases approaching $10~\mu$as. Importantly, they quantified how the parallax precision scales with angular separation between the target pulsar and the in-beam calibrator: for sufficiently bright targets observed eight times over approximately 18 months, per-epoch residuals correspond to parallax uncertainties of a few $\mu$as per arcminute of target-calibrator separation. This empirical scaling is essential for designing future, higher-precision VLBI campaigns.

More recently, \cite{2023MNRAS.519.4982D} presented the MSPSR$\pi$ catalogue, comprising VLBI astrometry for 18 MSPs and yielding statistically significant parallax distances for 15 of them. That work introduced refined calibration procedures (including ``inverse-referenced’’ interpolation) and a Bayesian astrometric pipeline capable of incorporating selected timing priors. The resulting analysis delivered percent-level distance constraints out to $\sim$1 kpc, measurements of transverse velocities, detections of angular broadening along some lines of sight, and updated constraints on dipolar gravitational radiation by combining acceleration and distance measurements.

The implications for PTA science are direct. VLBI measurements are approaching the $\sim 1$ pc absolute distance-precision regime for nearby, bright MSPs that already form the backbone of PTA data sets (e.g., J0437$-$4715, J0030+0451, J2145$-$0750). This trajectory aligns with the requirement established by \cite{PhysRevD.108.123535,2026PhRvD.113b2001K}: if even a small number of PTA pulsars achieve sub-parsec distance uncertainties, the localization of a nanohertz CW source can improve by orders of magnitude, enabling identification of a unique host galaxy. Consequently, VLBI astrometry of MSPs is transitioning from an auxiliary data product to an essential input to PTA-based GW source localization.

\section{SKA1-era Prospects for PTA-driven VLBI Astrometry}

SKA1-Mid will function as an exceptionally sensitive phased-array element within global VLBI networks. By coherently summing the signals from numerous SKA1-Mid antennas to form a tied-array beam, one obtains an effective station with extremely low system equivalent flux density. When this phased SKA1-Mid element is combined with long baselines to other instruments, such as the EVN (European VLBI Network), the uGMRT (upgraded Giant Metrewave Radio Telescope) and the FAST (Five-hundred-meter Aperture Spherical radio Telescope), the result is both high sensitivity to faint MSPs and sub-milliarcsecond angular resolution at $\sim$1.4 GHz \citep{2015aska.confE.143P,JIVE_D10_3_2019}. This is especially critical in the southern hemisphere, which hosts many of the most precisely timed and brightest PTA MSPs.

A key capability of SKA1-Mid for astrometry is its ability to form multiple simultaneous tied-array beams. This enables the pulsar and several nearby calibrators to be observed concurrently. Such an observing mode supports ``MultiView’’ astrometry \citep{2020A&ARv..28....6R}, in which phase solutions are derived from multiple calibrators surrounding the target rather than from a single reference source. MultiView astrometry explicitly models the spatially varying phase screen imposed by the ionosphere and troposphere, as well as residual structural variations in calibrator sources. This approach directly addresses the dominant systematics that typically limit sub-0.1 mas astrometry at GHz frequencies, and is therefore essential to approaching routine $\sim 10~\mu$as precision.

The connection to PTA science can be quantified through a representative forecast. Consider an observing program in the SKA1 era in which a bright PTA MSP with flux density $\sim 3$ mJy at 1.4 GHz is observed with SKA1-Mid and uGMRT in phased-array VLBI mode. Assume an on-source integration time of $\sim 3600$ s per epoch, repeated for $\sim 30$ epochs over $\sim$2-3 years. Under purely thermal noise assumptions (i.e.\ neglecting residual systematic errors), this strategy yields a parallax precision at the $\sim 10~\mu$as level. This corresponds to an absolute distance uncertainty of order $\sim 1$ pc for a pulsar at 300 pc, meeting the requirement for constraining the pulsar term phase.

\begin{table}[t]
\centering
\caption{Nearby PTA MSPs prioritized for a SKA1-MID + uGMRT VLBI parallax program. Columns list the pulsar name, distance, flux density at 1.4 GHz, and the expected distance precision (thermal-noise–limited estimates). Forecasts assume a phased SKA1-MID and phased uGMRT at 1.4 GHz, on-source integration of 3600 s per epoch, repeated over 30 epochs spanning 2-3 yrs with pulsar gating. The distance and flux density are taken from  the ATNF Pulsar Catalogue (http://www.atnf.csiro.au/research/pulsar/psrcat/) \citep{2005AJ....129.1993M}.}
\label{tab:MSPs}
\begin{tabular}{lccc}
\hline
name & distance & flux density @ 1.4GHz & expected precision\\
\hline
J0437-4715 & 157 pc & 150 mJy & 0.004 pc\\
J0030+0451 & 330 pc & 1.1 mJy & 2.57 pc\\
J1045-4509 & 340 pc & 2.7 mJy & 1.09 pc\\
J1744-1134 & 395 pc & 2.6 mJy & 1.55 pc\\
J2124-3358 & 410 pc & 4.5 mJy & 0.97 pc\\
J2145-0750 & 625 pc & 5.5 mJy & 1.84 pc\\
J1730-2304 & 650 pc & 4.0 mJy & 2.73 pc\\
\hline
\end{tabular}
\end{table}

The most relevant targets are bright, nearby PTA MSPs such as J0437-4715, J0030+0451, J1045-4509, J1744-1134, J2124-3358, J2145-0750, and J1730-2304 (Table~\ref{tab:MSPs}). Many of these pulsars already serve as cornerstone PTA clocks and are readily accessible to southern arrays. Crucially, \cite{2026PhRvD.113b2001K} show that if two or three MSPs with sub-parsec distance uncertainties are geometrically well placed in the vicinity of a CW source in the sky, the resulting PTA analysis can shrink the localization region for a strong signal from tens - hundreds of deg$^{2}$ to $\sim 10^{-3}$~deg$^{2}$ in favorable cases. As the number of MSPs with such precise, VLBI-quality distances increases, the sky area over which arcminute-scale CW localization is achievable will correspondingly expand, enlarging the domain of PTA sources that can be robustly associated with individual galaxies.

Sensitivity alone, however, is not sufficient. At $\sim$1 GHz, once thermal noise is suppressed, ionospheric and tropospheric fluctuations, as well as calibrator structural evolution, dominate the astrometric error budget. Classical single-calibrator phase referencing saturates at $\gtrsim 0.1$ mas when the angular separation between target and calibrator is large. The SKA-era approach, outlined in \cite{2015aska.confE.143P}, \cite{2020A&ARv..28....6R}, and \cite{JIVE_D10_3_2019}, is to design VLBI observations a priori to mitigate these systematic limitations. Specifically, SKA1-Mid’s multi-beam capability enables simultaneous observation of multiple calibrators and supports wideband, multi-frequency calibration. These capabilities underlie MultiView astrometry and are intended to deliver $\sim 10~\mu$as astrometric precision in practice, not merely in principle.

A strategic PTA-driven SKA1 VLBI program would include:
\begin{itemize}
\item Selection of a small set (5–10) of the brightest, most precisely timed PTA MSPs within $\sim$1 kpc, prioritizing those within $\sim 500$ pc.
\item Regular VLBI observations with SKA1-Mid and global baselines for $\sim$1 hr per epoch over $\sim$30 epochs spanning $\sim$2-3 years.
\item Use of multiple simultaneous beams to observe at least two nearby calibrators per pulsar for MultiView calibration, and the application of pulsar gating to boost signal-to-noise.
\item Joint astrometric analysis using Bayesian methods capable of incorporating timing-derived priors, as demonstrated by \cite{2023MNRAS.519.4982D}.
\item Injection of the resulting sub-parsec distance priors directly into PTA CW searches.
\end{itemize}

If implemented, such a program would provide PTAs with the ability to localize individual nanohertz CW sources.

Besides allowing us to better localize PTA CGW sources, the same astrometric data sets would have significant implications for fundamental pulsar and binary physics. High-precision VLBI astrometry improves constraints on relativistic binary parameters (e.g.\ via Shapiro delay and orbital geometry), refines measurements of transverse velocities and accelerations, constrains the Galactic electron density distribution, and enables tests of dipole gravitational radiation and possible time-variation of Newton’s constant \citep{2016ApJ...828....8D,2023MNRAS.519.4982D}. The SKA1-era astrometric program is therefore intrinsically multi-purpose: it supports GW source localization, compact-object astrophysics, and fundamental physics.

\section{Summary}

PTAs are approaching the regime in which they will detect individual continuous gravitational-wave sources from supermassive black hole binaries. To exploit such detections astrophysically, precise sky localization is essential: ideally at the arcminute level, so that the host galaxy can be uniquely identified. Conventional PTA CW searches typically yield localization regions spanning tens to hundreds of square degrees, which is insufficient for unique host association. The missing ingredient is an external, highly accurate measurement of pulsar distances.

VLBI astrometry of MSPs now routinely attains parallax precisions of a few tens of micro-arcseconds, and in favorable cases approaches $\sim 10~\mu$as. For nearby PTA MSPs (at distances of order 100–300 pc), this corresponds to absolute distance uncertainties of order 1 pc or better. This is precisely the threshold required to fix the pulsar term phase in PTA CW analyses. \cite{PhysRevD.108.123535,2026PhRvD.113b2001K} have demonstrated that, if even a small number of PTA pulsars have sub-parsec distance uncertainties, the localization area for a CW source can be reduced by orders of magnitude, reaching $\sim 10^{-3}$ deg$^2$ for favorable sky configurations. At that point, PTAs can associate a GW source with a unique host galaxy, obtain its redshift, and search for electromagnetic counterparts.

SKA1-Mid will render this capability practical on a systematic basis. Operating as a phased-array VLBI element with other large facilities (including uGMRT), SKA1-Mid will deliver both the sensitivity required to observe mJy-class MSPs and the multi-beam capability required for simultaneous multi-calibrator astrometry. Under realistic observing strategies (repeated $\sim$1 hr epochs over $\sim$2–3 years, pulsar gating, and MultiView calibration) forecasts indicate that SKA1-era VLBI will be able to measure distances to $\sim 1$ pc precision for the nearest PTA MSPs (e.g.\ J0437-4715, J0030+0451, J1045-4509, J1744-1134, J2124-3358). These distances can then be used as priors in PTA CW searches.

The scientific return of this synergy is immediate. Once a PTA can localize a nanohertz GW source to a region of a few arcminutes, it becomes feasible to identify the host galaxy, determine its redshift, infer physical (rather than purely strain-based) binary parameters, and search for electromagnetic counterparts such as dual active galactic nuclei, circumbinary gas, or periodic accretion signatures. This transforms PTA detections from anonymous GW signals into multi-messenger probes of specific SMBHBs in specific galactic environments. Simultaneously, the same VLBI astrometry improves pulsar timing models, enables tests of relativistic gravity in binary systems, and contributes to studies of Galactic structure.

\textbf{Acknowledgements}
KT is partially supported by JSPS KAKENHI Grant Numbers 20H00180, 21H01130, 21H04467, 24H01813, 26H00838, and 26K12345, and Bilateral Joint Research Projects of JSPS (120237710). HI is supported by the JSPS Bilateral Collaboration Program (ID:120239936) and the NINS International Collaborative Research Program (P.I. M.~Honma). HK is supported by JSPS KAKENHI Grant Number (25K00225). KN is supported by JSPS KAKENHI Grant Number (23H00118). FT is supported by the NAOJ Research Coordination Committee, NINS (NAOJ-RCC-24DS-0504).

\bibliographystyle{abbrvnat-maxbibnames4}
\bibliography{PTA-VLBI} 

\end{document}